\documentclass[preprint, number,12pt]{elsarticle2}

\usepackage[ansinew]{inputenc}
\usepackage[T1]{fontenc}
\usepackage{amsmath}
\usepackage{amssymb}
\usepackage[amsthm]{ntheorem}
\usepackage{scalefnt}
\usepackage{graphicx}
\usepackage{placeins}
\usepackage{pstricks}
\usepackage{pst-node}
\usepackage{mathrsfs}
\usepackage{bbm}

\usepackage{booktabs}
\usepackage{setspace}
\usepackage[scriptsize]{subfigure}

\newtheorem{satz}{Satz}
\newtheorem{Theorem}[satz]{Theorem}

\begin{document}

\begin{frontmatter}

\newtheorem{Corollary}{Corollary}
\newtheorem{Proposition}{Proposition}
\newtheorem{Lemma}{Lemma}
\newtheorem{Definition}{Definition}
\newtheorem{Example}{Example}
\newtheorem{Assumption}{Assumption}
\newtheorem{Remark}{Remark}
\renewcommand{\labelenumi}{(\roman{enumi})}
\newcommand{\sgn}{\operatorname{sgn}}

%------------------------------------------------------------------------------------------------------------------------
%*******************************************  TITLE  ********************************************************************
%------------------------------------------------------------------------------------------------------------------------
\title{Vine Constructions of Lévy Copulas}
%------------------------------------------------------------------------------------------------------------------------
%*******************************************  Authors  ******************************************************************
%------------------------------------------------------------------------------------------------------------------------
\author[lehrstuhl]{Oliver Grothe}
\ead{grothe@statistik.uni-koeln.de}
\author[lehrstuhl]{Stephan Nicklas\corref{cor1}}
\ead{nicklas@wiso.uni-koeln.de}

\cortext[cor1]{Corresponding author}

\address[lehrstuhl]{University of Cologne, Department of Economic and Social Statistics, Albertus-Magnus-Platz, 50923 Cologne, Germany}

%------------------------------------------------------------------------------------------------------------------------
%*******************************************   ABSTRACT   ***************************************************************
%------------------------------------------------------------------------------------------------------------------------
\begin{abstract}
\noindent L\'{e}vy copulas are the most general concept to capture jump dependence in multivariate L\'{e}vy processes.
They translate the intuition and many features of the copula concept into a time series setting. A challenge faced by both, distributional and L\'evy copulas, is to find flexible but still applicable models for higher dimensions. To overcome this problem, the concept of pair copula constructions has been successfully applied to distributional copulas. In this paper, we develop the pair construction for L\'{e}vy copulas (PLCC).
Similar to pair constructions of distributional copulas, the pair construction of a $d$-dimensional L\'{e}vy copula consists of $d(d-1)/2$ bivariate dependence functions. We show that only $d-1$ of these bivariate functions are L\'{e}vy copulas, whereas the remaining functions are distributional copulas. Since there are no restrictions concerning the choice of the copulas, the proposed pair construction adds the desired flexibility to L\'{e}vy copula models.
We discuss estimation and simulation in detail and apply the pair construction in a simulation study.
\end{abstract}


\end{frontmatter}
\doublespace

%----------------------------------------------------------------------------------------------------------------
%****************************************  SECTION 1: Introduction  *****************************************************
%------------------------------------------------------------------------------------------------------------------------
\section{Introduction}
%------------------------------------------------------------------------------------------------------------------------

Many financial and nonfinancial applications need multivariate models with jumps where the dependence of the jumps is captured adequately. To this end, Lévy processes have been applied in the literature. However, although the recently introduced concept of Lévy copulas enables modeling the dependence in Lévy processes in a multivariate setup, known parametric Lévy copulas are very inflexible in higher dimensions, i.e., they consist of very few parameters.
In this paper, we show that, similar to the pair copula construction of distributional copulas going back to Joe \cite{Joe1996}, Lévy copulas may be constructed from a constellation of parametric bivariate dependence functions. Because these dependence functions may be chosen arbitrarily, the resulting Lévy copulas flexibly capture various dependence structures.

Lévy processes are stochastic processes with independent increments. They consist of a Brownian motion part and jumps. Due to the jumps, Lévy processes capture stylized facts observed in financial data as non-normality, excessive skewness, and kurtosis (see, e.g., Johannes \cite{Johannes2004}). At the same time, they stay mathematically tractable and allow for derivative pricing by change of measure theory. For these reasons, intensive research is conducted on the statistical inference of Lévy processes (see, e.g., Lee and Hannig \cite{Lee2010} and the references therein).

The fundamental work for multivariate applications of Lévy processes is the seminal paper of Kallsen and Tankov \cite{KallsenTankov2004}, where the concept of Lévy copulas is introduced. This concept transfers the idea of distributional copulas to the context of Lévy processes. Distributional copulas (normally just referred to as \emph{copulas}) are functions which connect the marginal distribution functions of random variables to their joint distribution function. They contain the entire dependence information of the random variables (see, e.g., Nelsen \cite{Nelsen2006} for an introduction to copulas).
In the same sense, the theory of Lévy copulas enables to model multivariate Lévy processes by their marginal Lévy processes and to choose a suitable Lévy copula for the dependence structure separately. For papers regarding the estimation of Lévy copulas in multivariate Lévy processes and applications see, e.g., the recent papers of Esmaeili and Kl\"{u}ppelberg \cite{Esmaeili2010,Esmaeili2011,Esmaeili2011b} and references therein.

All papers involving Lévy copulas focus on rather small dimensions since higher-dimensional flexible Lévy copulas are difficult to construct. A similar effect has been observed during the first years of literature on distributional copulas, where mainly $2$-dimensional distributional copulas have been analyzed. One solution regarding distributional copulas has been the development of very flexible pair constructions of copulas going back to Joe \cite{Joe1996} and further developed in a series of papers (see, e.g., Bedford and Cooke \cite{Bedford2001} or Aas et al. \cite{Aas2009}). In pair copula constructions, a $d$-dimensional copula is constructed from $d(d-1)/2$ bivariate copulas. Here, $d-1$ of the bivariate copulas model the dependence of bivariate margins, whereas the remaining bivariate copulas model certain conditional distributions, such that the entire $d$-dimensional dependence structure is specified.

Lévy copulas are conceptually different from distributional copulas. While $d$-dimensional distributional copulas are distribution functions on a $[0,1]^d$ hypercube, $d$-dimensional Lévy copulas are defined on $\overline{\mathbb{R}}^d$ and relate to Radon measures. Therefore, the idea of pair constructions for copulas is not directly transferable to Lévy copulas and up to now it has not been clear whether it is possible at all. In this paper, we show that a pair copula construction of Lévy copulas (PLCC) is indeed possible. It also consists of $d(d-1)/2$ bivariate dependence functions but only $d-1$ of them are Lévy copulas, while the remaining ones are distributional copulas. For statistical inference, we derive sequential maximum likelihood estimators for an arbitrary pair construction of Lévy copulas as well as a simulation algorithm.
We analyze the applicability of the concept in a simulation study. The estimation and simulation algorithms show encouraging results in a finite sample setting.

The remainder of the paper is structured as follows. In Section \ref{sec.Cop_uPCop}, we review the theory of copulas for random variables and pair copula constructions of such copulas. In Section \ref{sec.LevProcLevCop}, we address the theory of Lévy processes and Lévy copulas. Our pair construction of Lévy copulas is derived in Section \ref{sec.PaLevCop}. In Section \ref{sec.SimEst}, we provide simulation as well as maximum likelihood estimation methods. Section \ref{sec.Simu} contains simulation studies probing the simulation and estimation algorithms in finite samples and Section \ref{sec.Concl} concludes.

%------------------------------------------------------------------------------------------------------------------------
\section{Preliminaries}
%------------------------------------------------------------------------------------------------------------------------
In this section, we briefly recall necessary theory on copulas, pair copulas, Lévy processes, and the Lévy copula concept.

\subsection{Copulas and Pair Copula Construction}
\label{sec.Cop_uPCop}

Let $X=(X_1,\dots,X_d)$ be a random vector with joint distribution function $F$ and continuous marginal distribution functions $F_i$, $i=1,\dots,d$. The copula $C$ of $X$ is the uniquely defined distribution function with domain $[0,1]^d$ and uniformly distributed margins satisfying
\begin{equation*}
F(x_1,\dots,x_d)=C\big[F_1(x_1),\dots,F_d(x_d)\big].
\end{equation*}
By coupling the marginal distribution functions to the joint one, the copula $C$ entirely determines the dependence of the random variables $X_1,\dots,X_d.$ While many $2$-dimensional parametric families of copulas exist, see, e.g., Nelsen \cite{Nelsen2006}, the families for the $d$-dimensional case suffer from lack of flexibility. To overcome this problem, the concept of pair copula construction has been developed (see, e.g., Joe \cite{Joe1996} for the seminal work or the detailed introductions in Aas et al. \cite{Aas2009}, Bedford and Cooke \cite{Bedford2001}, and Berg and Aas \cite{Berg2009}). In a pair copula construction, a $d$-dimensional copula $C(u_1,\dots,u_d)$ is constructed of $d(d-1)/2$ bivariate copulas. Of these bivariate copulas, $d-1$ bivariate copulas directly model $d-1$ $2$-dimensional margins of the copula $C$, whereas the other bivariate copulas indirectly specify the remaining parts in terms of conditional distributions.
Since the number of possible combinations grows rapidly with the dimension, Bedford and Cooke \cite{Bedford2001, Bedford2002} introduced a graphical model, called regular vines (R-vines), to describe the structures of pair copula constructions.

\begin{figure}
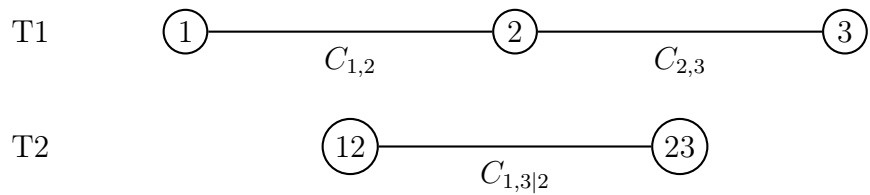

        \begin{center}
        \begin{psmatrix}[colsep=1.5cm,rowsep=0.8cm]
        T1&\circlenode{1}{1} && \circlenode{2}{2} && \circlenode{3}{3}\\
        T2&&\circlenode{12}{12}&&\circlenode{23}{23}
        \ncline{2}{1}\naput{$ {C}_{1,2}$}
        \ncline{3}{2}\naput{$ {C}_{2,3}$}

        \ncline{23}{12}\naput{$C_{1,3|2}$}
        \end{psmatrix}
        \end{center}
        \caption{\label{fig.vine} Example of a pair copula construction of a $3$-dimensional copula. It consists of 2 trees and 3(3-1)/2=3 bivariate copulas $ {C}_{1,2}$, $ {C}_{2,3}$ and $C_{1,3|2}$.}
\end{figure}

An example of a regular vine for the 3-dimensional case is given in Figure \ref{fig.vine}. It shows three dimensions (labeled 1,2 and 3) and two trees (labeled T1 and T2) of dependence functions. The first tree (T1) contains the two bivariate copulas, $C_{1,2}$ and $C_{2,3},$ modeling the dependence between dimensions 1 and 2 and dimensions 2 and 3, respectively. Thus, tree T1 completely determines these two bivariate dependence structures. It also indirectly determines parts of the dependence between dimensions 1 and 3, but not necessarily the entire dependence. For instance, if the pairs 1,2 and 2,3 are each correlated with 0.9, then 1 and 3 cannot be independent but their exact dependence is not specified. In particular, the conditional dependence of 1 and 3 given 2 is not specified. Therefore, in the second tree (T2), this bivariate conditional dependence is modeled with another copula, $C_{1,3|2}.$ Together, the three bivariate copulas fully specify the dependence of the three dimensions. Since the choice of all three bivariate copulas is arbitrary, the vine structure provides a very flexible way to construct multidimensional copulas. In the $d$-dimensional case, $d(d-1)/2$ bivariate copulas are needed and arranged in $d-1$ trees (see, e.g., Joe \cite{Joe1996}). There are special cases of regular vines, e.g., C-vines or D-vines (see, e.g., Aas et al. \cite{Aas2009} for a more detailed introduction).

%------------------------------------------------------------------------------------------------------------------------
\subsection{Lévy Processes and Lévy Copulas}
\label{sec.LevProcLevCop}

Detailed information about Lévy processes may be found in Rosinski \cite{Rosinski2001}, Kallenberg \cite{Kallenberg2002} or Sato \cite{Sato1999}. Introductions to Lévy copulas are given in Kallsen and Tankov \cite{KallsenTankov2004} or Cont and Tankov \cite{ContTankov}. Here, we give a very short overview of both.

Let $(\Omega, \mathscr{F},P)$ be a probability space. A Lévy process $(L_t)_{t \in \mathbb{R}_+}$ is a stochastic process with stationary, independent increments starting at zero.
Lévy processes can be decomposed into a deterministic drift function, a Brownian motion part and a pure jump process with a possibly infinite number of small jumps, see, e.g., Kallenberg \cite{Kallenberg2002}, Theorem 15.4 (Lévy It\^{o} decomposition). In this paper, we focus on spectrally positive Lévy processes, which are Lévy processes with positive jumps only. This facilitates the notation considerably and in many relevant cases it is sufficient to consider positive jumps only. However, all results of the paper may be extended to the general case.
The characteristic function of the distribution of such an $\mathbb{R}^d$-valued spectrally positive Lévy process $L_t$, at time $t$, is given by the Lévy-Khinchin representation (see Kallenberg \cite{Kallenberg2002})
\begin{equation}
\varphi_{L_t}(z)=\exp\left\{t\left(i\langle \gamma,z\rangle  - \frac{1}{2}\langle z, \Sigma z \rangle +\int_{\mathbb{R}^d_+}(e^{i\langle z,x\rangle}-1)\nu(dx)
\right)\right\}.
\label{EquationLevyKhintchineSubordinator}
\end{equation}

Here, $\gamma\in \mathbb{R}^d$ corresponds to the drift part of the process and $\Sigma$ is the covariance matrix of the Brownian motion part at time $t=1.$ The Lévy measure $\nu$ is a measure on $\mathbb{R}^d$ which is concentrated on the positive domain $\mathbb{R}^d_+\setminus \{0\}$ with $\int_{\mathbb{R}^d}x\nu(dx)<\infty.$
The Lévy measure completely characterizes the jump parts of the Lévy process, where $\nu(A)$ for $A\in \mathcal{B}(\mathbb{R}^d_+)$ is the expected number of jumps per unit of time with jump sizes in $A.$ A spectrally positive Lévy process with positive entrees of $\gamma$ and $\Sigma=0$ is called subordinator. It has no negative increments.

An interesting example for a one-dimensional subordinator is the stable subordinator. It is heavy tailed and therefore suggested as a loss process for operational risk models.
In Basawa and Brockwell \cite{Basawa1978},
the Lévy measure of a stable subordinator on $\mathbb{R}_+$ is defined by
\begin{equation*}
\nu(B)=\int_{\mathbb{R}_+}\mathbbmss{1}_B(z)\frac{\alpha\beta}{z^{\alpha+1}}dz,
\end{equation*}
where $\alpha \in (0,1)$ and $\beta>0$.

Related to the Lévy measure, its tail integral is defined by (see, e.g., Definition 3.1 in Esmaeili and Kl\"{u}ppelberg \cite{Esmaeili2010})
\begin{equation*}
U(x_1,\ldots,x_d)=
\begin{cases}
\nu([x_1,\infty)\times\ldots\times [x_d,\infty))&\text{if }(x_1,\ldots,x_d)\in [0,\infty)^d\backslash \{0\},\\
0 & \text{if } x_i=\infty \text{ for at least one $i$},\\
\infty & \text{if } (x_1,\ldots,x_d)=0.
\end{cases}
\end{equation*}
The tail integral $U$ of a spectrally positive Lévy process uniquely determines its Lévy measure $\nu$.
We define the marginal tail integrals $U_k$ for any dimension $k=1,\ldots,d$ of the multivariate Lévy process in a similar way.
For one-dimensional spectrally positive Levy measures $\nu,$ the tail integral is $U(x)=\nu([x,\infty)),$ i.e., the expected number of jumps  per unit of time with jump sizes larger or equal to $x.$
For the one-dimensional stable subordinator, the tail integral can be explicitly calculated and inverted for $x,u>0$,
\begin{equation*}
U(x)=\int_{[x,\infty)}\frac{\alpha\beta}{z^{\alpha+1}}dz=\beta x^{-\alpha}\,\,\,\text{with}\,\,\,U^{-1}(u)=\left(\frac{u}{\beta}\right)^{-\frac{1}{\alpha}}.
\end{equation*}
The inverse of the tail integral is needed for the simulation of the process.

Dependence of jumps of a multivariate Lévy process can be described by a Lévy copula which couples the marginal tail integrals to the joint one.
A $d$-dimensional Lévy copula is a measure defining function $\mathfrak{C}(u_1,\dots,u_d):[0,\infty]^d\rightarrow [0,\infty]$ with margins $\mathfrak{C}_k(u_k):=\mathfrak{C}(\infty,\dots,\infty,u_k,\infty,\dots,\infty)=u_k$ for all $u_k\in[0,\infty]$ and $k=1,\dots,d.$
In particular, let $U$ denote the tail integral of a spectrally positive $d$-dimensional Lévy process whose components have the tail integrals $U_1,\ldots,U_d$. Then, there exists a Lévy copula $\mathfrak{C}$ such that for all $(x_1,\ldots,x_d)\in \overline{\mathbb{R}}^d_+$
\begin{equation}
U(x_1,\ldots,x_d)=\mathfrak{C}(U_1(x_1),\ldots,U_d(x_d)).
\label{EquationSklarLevyCopulas}
\end{equation}
Conversely, if $\mathfrak{C}$ is a Lévy copula and $U_1,\ldots,U_d$ are marginal tail integrals of spectrally positive Lévy processes, Equation \eqref{EquationSklarLevyCopulas} defines the tail integral of a $d$-dimensional spectrally positive Lévy process and $U_1,\ldots,U_d$ are the tail integrals of its components. Both statements are often called the Sklar's theorem for Lévy copulas and are proved, e.g, in Cont and Tankov \cite{ContTankov}.

In this paper, we focus on Lévy copulas for which the following assumption holds.
\begin{Assumption}  \emph{} \\
\label{AssumptionLimes}
Let $\mathfrak{C}_{1,\ldots,d}$ be a L\'evy copula such that for every $I \subset \{1,\ldots,d\}$ nonempty,
\begin{equation}
\lim_{(u_i)_{i\in I}\rightarrow \infty}\mathfrak{C}_{1,\ldots, d}(u_1,\ldots,u_d)=\mathfrak{C}_{1,\ldots,d}(u_1,\ldots,u_d)|_{(u_i)_{i\in I}=\infty}.
\end{equation}
\end{Assumption}
This is a rather weak assumption on the Lévy copula and is assumed in many papers, e.g., in Tankov \cite{Tankov2005}.
It means that the Lévy copula has no new information at the points $u_i=\infty$ which is not already contained in the limit for $u_i\rightarrow \infty.$ We need it since it ensures a bijection between a Lévy copula on $\overline{\mathbb{R}}^d_+$ and a positive measure $\mu_{1,\ldots,d}$ on $\mathcal{B}(\mathbb{R}^d_+)$ with one-dimensional Lebesgue margins. This measure is given by
\begin{equation}
\label{EquationMeasureByLC}
\mu_{1,\ldots,d}((a,b])=V_{\mathfrak{C}_{1,\ldots,d}}([a,b]),
\end{equation}
where $a,b \in \mathbb{R}^d_+$ with $a\leq b$, component-wise, and $V_{\mathfrak{C}_{1,\ldots,d}}$ refers to the ${\mathfrak{C}_{1,\ldots,d}}$-volume of the $d$-box $[a,b]$ which is defined as $$V_{\mathfrak{C}_{1,\ldots,d}}([a,b])=\sum \text{sgn}(c)\mathfrak{C}_{1,\ldots,d}(c).$$ The sum is taken over all vertices $c$ of $[a,b]$ and $$\text{sgn}(c)=\begin{cases}1& \text{if $c_k=a_k$ for an even number of k}, \\ -1& \text{if $c_k=a_k$ for an odd number of k}.  \end{cases}$$

Furthermore, any positive measure $\mu_{1,\ldots,d}$ on $\mathbb{R}^d_+$ with Lebesgue margins uniquely defines a Lévy copula on $\overline{\mathbb{R}}^d_+$ that satisfies Assumption \ref{AssumptionLimes} by
\begin{equation*}
\mathfrak{C}_{1,\ldots,d}(u_1,\ldots,u_d):=\mu_{1,\ldots,d}([0,u_1]\times \ldots \times[0,u_d])
\end{equation*}
and by setting
\begin{equation*}
\mathfrak{C}_{1,\ldots,d}(u_1,\ldots,u_d)|_{(u_i)_{i\in I}=\infty}:=\lim_{(u_i)_{i\in I}\rightarrow \infty}\mu_{1,\ldots,d}([0,u_1]\times \ldots \times[0,u_d]).
\end{equation*} These results are proved, e.g., in Section 4.5 in Kingman and Taylor \cite{Kingman1966}.

An example for a Lévy copula which is used later in the paper is the Clayton Lévy copula. For spectrally positive, 2-dimensional Lévy processes it is given on $\mathbb{R}_+$ by
\begin{align}\label{glg.clayton}\mathfrak{C}(u,v)=\left(u^{-\theta}+v^{-\theta}\right)^{-1/\theta}.\end{align}
Here, $\theta>0$ determines the dependence of the jump sizes, where larger values of $\theta$ indicate a stronger dependence.

%------------------------------------------------------------------------------------------------------------------------
\section{Pair Lévy Copulas}
%------------------------------------------------------------------------------------------------------------------------
\label{sec.PaLevCop}
In this section, we present the pair construction of $d$-dimensional Lévy copulas. In particular, we show that analogously to the pair construction of distributional copulas, $d(d-1)/2$ functions of bivariate dependence may be arranged such that they define a $d$-dimensional Lévy copula. In Sections \ref{Example3D} and \ref{Example4D}, we provide illustrating examples how to construct multivariate pair Lévy copula constructions. Readers not interested in the technical parts may read these examples first.

\subsection{Technical Part}

The central theorem for the construction is Theorem \ref{TheoremPLCC}. It states that two $(d-1)$-dimensional Lévy copulas with overlapping $(d-2)$-dimensional margins may be coupled to an $d$-dimensional Lévy copula by a new, $2$-dimensional distributional copula. Ensured by vine constructions (see Bedford and Cooke \cite{Bedford2002}) and starting at $(d-1)=2,$ Theorem \ref{TheoremPLCC} therefore enables to sequentially construct Lévy copulas out of $2$-dimensional dependence functions, i.e., $2$-dimensional distributional copulas and Lévy copulas.
Before we state the theorem, for convenience, we recall some definitions which can be found, e.g., in  Ambrosio et al. \cite{Ambrosio2000}.

\begin{Definition} \emph{} \\
A positive measure on $(\mathbb{R}^d_+,\mathcal{B}(\mathbb{R}^d_+))$ that is finite on compact sets is called a {\em positive Radon measure}.\\
Let $(X,\mathcal{E})$ and $(Y,\mathcal{F})$ be measure spaces and let $f:X\rightarrow Y$ be a measurable function. For any measure $\mu$ on $(X,\mathcal{E})$, we define the {\em Push Forward Measure} $f_{\#}\mu$ in $(Y,\mathcal{F})$ by
\begin{equation*}
f_{\#}\mu:=\mu\left(f^{-1}(K)\right)\quad \forall K\in \mathcal{F}.
\end{equation*}
Let $\mu$ be a positive Radon measure on $\mathbb{R}^d_+$ and $x\mapsto \xi_x$ a function which assigns a finite Radon measure $\xi_x$ on $\mathbb{R}^m_+$ to each $x\in \mathbb{R}^d_+$. We say this {\em map is $\mu$-measurable} if $x\mapsto \xi_x(B)$ is $\mu$-measurable for any $B\in \mathcal{B}(\mathbb{R}^m_+)$.
\end{Definition}

\begin{Definition}[Generalized Product] \emph{} \\
Let $\mu$ be a positive Radon measure on $\mathbb{R}^d_+$ and $x\mapsto \xi_x$ a $\mu$-measurable function which assigns a probability measure $\xi_x$ on $\mathbb{R}^m_+$ to each $x\in \mathbb{R}^d_+$. We denote by $\mu \otimes \xi_x$ the Radon measure on $\mathbb{R}^{d+m}_+$ defined by
\begin{equation*}
\mu \otimes \xi_x(B):=\int_{\mathbb{R}^d_+}\left(\int_{\mathbb{R}^m_+}\mathbbmss{1}_B(x,y)d\xi_x(y)\right)d\mu(x) \quad \forall B\in\mathcal{B}(K\times \mathbb{R}^m_+),
\end{equation*}
where $K\subset \mathbb{R}^d_+$ is any compact set.
\end{Definition}

We also need a theorem which states that a Radon measure may be decomposed into a a projection onto some of its dimensions and a probability measure. For a proof see Theorem 2.28 in Ambrosio et al. \cite{Ambrosio2000} and also the sentence after Corollary 2.29 there.

\begin{Theorem}[Disintegration] \emph{}\label{TheoremDisintefration} \\
Let $\mu_{1,\ldots,d+m}$ be a Radon measure on $\mathbb{R}^{d+m}_+$, $\pi: \mathbb{R}^{d+m}_+\mapsto \mathbb{R}^d_+$ the projection on the first $d$ variables and $\mu_{1,\ldots,d}=\pi_\#\mu_{1,\ldots,d+m}$. Let us assume that $\mu_{1,\ldots,d}$ is a positive Radon measure, i.e., that $\mu_{1,\ldots,d+m}(K\times \mathbb{R}^m_+)<\infty$ for any compact set $K\subset \mathbb{R}^d_+$. Then, there exists a finite measure $\xi_x$ in $\mathbb{R}^m_+$ such that $x\mapsto \xi_x$ is $\mu_{1,\ldots,d}$-measurable, $\xi_x$ is a probability measure almost everywhere in $\mathbb{R}^d_+$, and
\begin{equation*}
\int_{\mathbb{R}^{d+m}_+}\mathbbmss{1}_B(x,y)d\mu_{1,\ldots,d+m}(x,y)=\int_{\mathbb{R}^d_+}\left( \int_{\mathbb{R}^m_+} \mathbbmss{1}_B(x,y)\xi_x(y) \right)d\mu_{1,\ldots,d}(x),
\end{equation*}
this is $\mu_{1,\ldots,d+m}(B)=\mu_{1,\ldots,d}\otimes\xi_x(B)$ for any $B\in\mathcal{B}(K\times \mathbb{R}^m_+),$ where $K\subset \mathbb{R}^d_+$ is any compact set.
\end{Theorem}

We are now able to state the main theorem.

\begin{Theorem}[Pair Lévy Copula Composition]  \emph{} \\
Let $\mathfrak{C}_{1,\ldots,d-1}$ and $\mathfrak{C}_{2,\ldots,d}$ be two Lévy copulas on $\overline{\mathbb{R}}_+^{d-1}$ where $\mathfrak{C}_{1,\ldots,d-1}$ is a Lévy copula on the variables $u_1,\ldots,u_{d-1}$ and $\mathfrak{C}_{2,\ldots,d}$ is a Lévy copula on the variables $u_2,\ldots,u_{d}$. Denote the corresponding measures on $\mathbb{R}_+^{d-1}$ by $\mu_{1,\ldots,d-1}$ and $\mu_{2,\ldots,d}$, respectively. Suppose that the two measures have an identical $(d-2)$-dimensional margin $\mu_{2,\ldots,d-1}$ on the variables $u_2,\ldots,u_{d-1}$.
Then, we can define a Lévy copula on $\mathbb{R}^d_+$ by
\begin{equation*}
\mathfrak{C}_{1,\ldots,d}(u_1,\ldots,u_d):=\int\limits_{[0,u_2]\times\ldots\times [0,u_{d-1}]}C(F_{1|z_2,\ldots,z_{d-1}}(u_1),F_{d|z_2,\ldots,z_{d-1}}(u_d))d\mu_{2,\ldots ,d-1}(z_2,\ldots,z_{d-1}),
\end{equation*}
where $F_{1|u_2,\ldots,u_{d-1}}$ is the one-dimensional distribution function corresponding to the probability measure $\xi_{1|u_2,\ldots,u_{d-1}}$ from the decomposition of $\mu_{1,\ldots,d-1}$ into
\begin{equation*}
\mu_{1,\ldots,d-1}=\mu_{2,\ldots,d-1} \otimes \xi_{1|u_2,\ldots,u_{d-1}},
\end{equation*}
$F_{d|u_2,\ldots,u_{d-1}}$ is the one-dimensional distribution function corresponding to the probability measure $\xi_{d|u_2,\ldots,u_{d-1}}$ from the decomposition of $\mu_{2,\ldots,d}$ into
\begin{equation*}
\mu_{2,\ldots,d}=\mu_{2,\ldots,d-1} \otimes \xi_{d|u_2,\ldots,u_{d-1}},
\end{equation*}
and $C$ is a distributional copula. Since Lévy copulas are functions on $\overline{\mathbb{R}}^d_+$, we set for every $I\subset\{1,\ldots,d\}$ nonempty,
\begin{equation}
\mathfrak{C}_{1,\ldots,d}(u_1,\ldots,u_d)|_{(u_i)_{i\in I}=\infty}:=\lim_{(u_i)_{i\in I}\rightarrow \infty}\mathfrak{C}_{1,\ldots,d}(u_1,\ldots,u_d).
\label{EqPLCCLimes}
\end{equation}
\label{TheoremPLCC}
\end{Theorem}

The theorem, which is proved in the appendix, illustrates how to construct a $d$-dimensional Lévy copula from two $(d-1)$-dimensional Lévy copulas with a common margin. Applying the theorem recursively, these $(d-1)$-dimensional Lévy copulas can be constructed from $(d-2)$-dimensional ones. This can be repeated down to construct $3$-dimensional Lévy copulas from bivariate ones.
In higher dimensions, there are many ways for this procedure due to possible permutations of the dimensions and numerous possible pairwise combinations within the trees.
 The graphical visualization of the different resulting structures of pair construction is possibly by the concept of regular vines as developed in Bedford and Cooke \cite{Bedford2002}. Regular vines also help to construct pair Lévy copulas top-down. This means to start with $d-1$ bivariate Lévy copulas and to combine them successively to $3, 4, 5 ,\dots, d$-dimensional Lévy copulas.
 The regular vine approach ensures that at each step the involved Lévy copulas have sufficiently overlapping margins, and that therefore the theorem can be applied.
To illustrate this procedure, we give two detailed examples. The first example refers to the most simple case, a $3$-dimensional Lévy copula. The second, 4-dimensional example then illustrates how to sequentially add dimensions to the pair copula construction.

\subsection{Example: 3-dimensional Pair Lévy Copula Construction} \label{Example3D} 
\begin{figure}
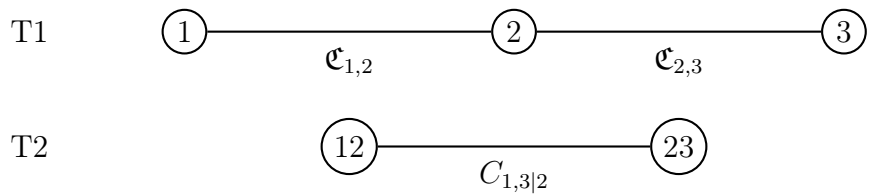

        \begin{center}
        \begin{psmatrix}[colsep=1.5cm,rowsep=0.8cm]
        T1&\circlenode{1}{1} && \circlenode{2}{2} && \circlenode{3}{3}\\
        T2&&\circlenode{12}{12}&&\circlenode{23}{23}
        \ncline{2}{1}\naput{$\mathfrak{C}_{1,2}$}
        \ncline{3}{2}\naput{$\mathfrak{C}_{2,3}$}

        \ncline{23}{12}\naput{$C_{1,3|2}$}
        \end{psmatrix}
        \end{center}
        \caption{\label{fig.levpair1} Pair construction of a 3-dimensional Lévy copula out of $3(3-1)/2=3$ bivariate dependence functions. The functions $\mathfrak{C}_{1,2}$ and $\mathfrak{C}_{2,3}$ in the first tree are Lévy copulas, while $C_{1,3|2}$ in the second tree is a distributional copula.}
\end{figure}

A 3-dimensional example can be constructed applying Theorem \ref{TheoremPLCC} to combine two 2-dimensional Lévy copulas by a distributional copula.
As in the usual pair copula construction for distributional copulas, in Figure \ref{fig.levpair1} we use the vine concept to visualize the resulting dependence structure.
The bivariate dependence structures in the first tree are Lévy copulas, whereas the copula in the second tree is a distributional copula.
From Theorem \ref{TheoremPLCC} follows that
\begin{equation*}
\mathfrak{C}_{1,2,3}(u_1,u_2,u_3)
=\int\limits_{[0,u_2]}C_{1,3|2}(F_{1|z_2}(u_1),F_{3|z_2}(u_3))d\mu_2(z_2)
\end{equation*}
is a Lévy copula, where $F_{1|u_2}(u_1)$ is the one-dimensional distribution function corresponding to the probability measure $\xi_{1|u_2}$ from the decomposition of $\mu_{1,2}$ into
\begin{equation}
\mu_{1,2}=\mu_{2} \otimes \xi_{1|u_2}
\label{Equation3DVine1}
\end{equation}
and $F_{3|u_2}$ is the one-dimensional distribution function corresponding to the probability measure $\xi_{3|u_2}$ from the decomposition of $\mu_{2,3}$ into
\begin{equation}
\mu_{2,3}=\mu_{2} \otimes \xi_{3|u_2}.
\label{Equation3DVine2}
\end{equation}

Remember that $\mu_{1,2}$ is the Radon measure corresponding to $\mathfrak{C}_{1,2}.$ With Theorem \ref{TheoremDisintefration} and the considerations after Assumption \ref{AssumptionLimes} we see that $\mu_2$ in Equation \eqref{Equation3DVine1} is the Lebesgue measure. Analogously, $\mu_{2,3}$ is the Radon measure corresponding to $\mathfrak{C}_{2,3}$ and therefore $\mu_2$ in Equation \eqref{Equation3DVine2} is the Lebesgue measure as well.

To check whether $\mathfrak{C}_{1,2,3}(u_1,u_2,u_3)$ has the desired margins, we calculate
\begin{eqnarray*}
\mathfrak{C}_{1,2,3}(u_1,u_2,\infty)&=&\int\limits_{[0,u_2]}C_{1,3|2}(F_{1|z_2}(u_1),F_{3|z_2}(\infty))dz_2\\
&=&\int\limits_{[0,u_2]}C_{1,3|2}(F_{1|z_2}(u_1),1)dz_2\\
&=&\int\limits_{[0,u_2]}F_{1|z_2}(u_1)dz_2\\
&=&\int\limits_{[0,u_2]}\left(\int\limits_{[0,u_1]}d\xi_{1|z_2}(z_1)\right)dz_2\\
&=&\int\limits_{[0,u_1]\times[0,u_2]}d\mu_{1,2}(z_1,z_2)\\
&=&\mathfrak{C}_{1,2}(u_1,u_2).
\end{eqnarray*}
A similar procedure shows that
\begin{equation*}
\mathfrak{C}_{1,2,3}(\infty,u_2,u_3)=\mathfrak{C}_{2,3}(u_2,u_3).
\end{equation*}
As expected, we do not get such a direct representation of the third bivariate margin
\begin{equation*}
\mathfrak{C}_{1,2,3}(u_1,\infty,u_3)=\int\limits_{[0,\infty)}C_{1,3|2}(F_{1|z_2}(u_1),F_{3|z_2}(u_3))dz_2
\end{equation*}
because this margin is not only influenced by the distributional copula $C_{1,3|2}$ but also by $\mathfrak{C}_{1,2}$ and $\mathfrak{C}_{2,3}$. However, we can adjust the bivariate margin of the first and third dimension by changing $C_{1,3|2}$ without affecting the other two bivariate margins.

\subsection{Example: 4-dimensional Pair Lévy Copula Construction} \label{Example4D}
\begin{figure}
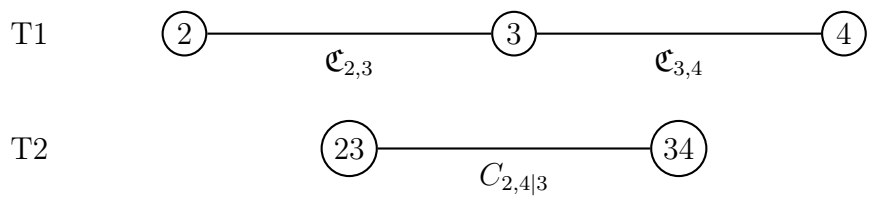

        \begin{center}
        \begin{psmatrix}[colsep=1.5cm,rowsep=0.8cm]
        T1&\circlenode{2}{2} && \circlenode{3}{3} && \circlenode{4}{4}\\
        T2&&\circlenode{23}{23}&&\circlenode{34}{34}
        \ncline{3}{2}\naput{$\mathfrak{C}_{2,3}$}
        \ncline{4}{3}\naput{$\mathfrak{C}_{3,4}$}

        \ncline{34}{23}\naput{$C_{2,4|3}$}
        \end{psmatrix}
        \end{center}
        \caption{\label{fig.levcop3_2} Pair construction of the second three dimensions of a 4-dimensional Lévy copula out of $3(3-1)/2=3$ bivariate dependence functions. The functions $\mathfrak{C}_{2,3}$ and $\mathfrak{C}_{3,4}$ in the first tree are Lévy copulas, while $C_{2,4|3}$ in the second tree is a distributional copula. The Lévy copula $\mathfrak{C}_{2,3}$ is the same Lévy copula as in Figure \ref{fig.levpair1} which refers to a the pair construction of the first three dimensions.}
\end{figure}

Considering $4$ dimensions, we need two $3$-dimensional Lévy copulas with an identical $2$-dimensional margin. Here, we reuse the Lévy copula from Example \ref{Example3D} for the first three dimensions. The second $3$-dimensional Lévy copula is constructed in the same way and has the vine representation shown in Figure \ref{fig.levcop3_2}.

\begin{figure}
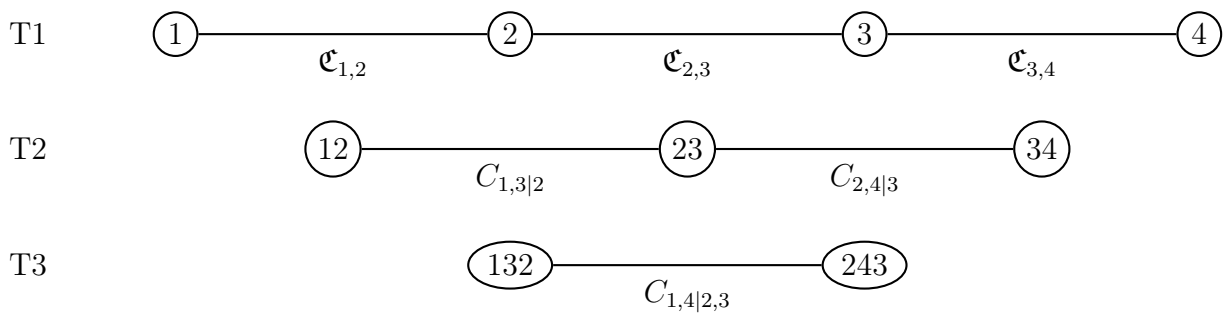

        \begin{center}
        \begin{psmatrix}[colsep=1.4cm,rowsep=0.8cm]
        T1&\circlenode{1}{1}&&\circlenode{2}{2} && \circlenode{3}{3} && \circlenode{4}{4}\\
        T2&&\circlenode{12}{12}&&\circlenode{23}{23}&&\circlenode{34}{34}\\
        T3&&&\ovalnode{132}{132}&&\ovalnode{243}{243}
        \ncline{2}{1}\naput{$\mathfrak{C}_{1,2}$}
        \ncline{3}{2}\naput{$\mathfrak{C}_{2,3}$}
        \ncline{4}{3}\naput{$\mathfrak{C}_{3,4}$}

        \ncline{23}{12}\naput{$C_{1,3|2}$}
        \ncline{34}{23}\naput{$C_{2,4|3}$}

        \ncline{243}{132}\naput{$C_{1,4|2,3}$}
        \end{psmatrix}
        \end{center}
             \caption{\label{fig.levcop4_1} Combination of the first three dimensions and the second three dimensions to a pair construction of a 4-dimensional Lévy copula. It consists of $4(4-1)/2=6$ bivariate dependence functions. Only the functions in the first tree are Lévy copulas, while the functions in the second and third tree are distributional copulas. }
\end{figure}

Notice that the Lévy copula $\mathfrak{C}_{2,3}$ is used in both $3$-dimensional pair Lévy copulas. Therefore, the marginal Lévy copulas
\begin{equation*}
\mathfrak{C}_{1,2,3}(\infty,u_2,u_3)=\mathfrak{C}_{2,3}(u_2,u_3)=\mathfrak{C}_{2,3,4}(u_2,u_3,\infty)
\end{equation*}
are the same and we can apply Theorem \ref{TheoremPLCC} to construct a $4$-dimensional Lévy copula with the vine representation shown in Figure \ref{fig.levcop4_1}
and
\begin{equation*}
\mathfrak{C}_{1,2,3,4}(u_1,u_2,u_3,u_4)
=\int\limits_{[0,u_2]\times[0,u_3]}C_{1,4|2,3}(F_{1|z_2,z_3}(u_1),F_{4|z_2,z_3}(u_4))d\mu_{2,3}(z_2,z_3)
\end{equation*}
where $F_{1|u_2,u_3}$ is the one-dimensional distribution function corresponding to the probability measure $\xi_{1|u_2,u_3}$ from the decomposition of $\mu_{1,2,3}$ from the first pair Lévy copula $\mathfrak{C}_{1,2,3}$ into
\begin{equation*}
\mu_{1,2,3}=\mu_{2,3} \otimes \xi_{1|u_2,u_3}.
\end{equation*}
The one-dimensional distribution function $F_{4|u_2,u_3}$ corresponds to the probability measure $\xi_{4|u_2,u_3}$ from the decomposition of $\mu_{2,3,4}$ from the second pair Lévy copula $\mathfrak{C}_{2,3,4}$ into
\begin{equation*}
\mu_{2,3,4}=\mu_{2,3} \otimes \xi_{4|u_2,u_3}.
\end{equation*}

%------------------------------------------------------------------------------------------------------------------------
\section{Simulation and Estimation}
%------------------------------------------------------------------------------------------------------------------------
\label{sec.SimEst}

In this section we discuss the simulation of multivariate Lévy processes as well as the maximum likelihood estimation of the pair Lévy copula. We need the following assumption which is fulfilled by the common parametric families of the bivariate (Lévy) copulas.
\begin{Assumption} \emph{} \\
In the following, we assume that all bivariate distributional and Lévy copulas are continuously differentiable.
\label{AssumptionContDiff}
\end{Assumption}

\subsection{Simulation}

The simulation of multivariate Lévy processes built upon Lévy copulas bases on a series representation for Lévy processes and the following theorem.

\begin{Theorem} \emph{} \\
Let $\nu$ be a Lévy measure on $\mathbb{R}^d_+$, satisfying $\int_{\mathbb{R}^d_+}(\Vert x\Vert \wedge 1) d\nu(x)<\infty$, with marginal tail integrals $U_i$, $i=1,\ldots,d$, Lévy copula $\mathfrak{C}_{1,\ldots,d}$ with corresponding measure $\mu_{1,\dots,d}$. Let $(V_i)_{i\in \mathbb{N}}$ be a sequence of independent and uniformly $[0,1]$ distributed random variables and $(\Gamma^1_i,\ldots,\Gamma^{d-1}_i)_{i\in \mathbb{N}}$ be a Poisson point process on $\mathbb{R}^{d-1}_+$ with intensity measure $\mu_{1,\ldots,d-1}$ from the decomposition of
\begin{equation*}
\mu_{1,\ldots,d}=\mu_{1,\ldots,d-1}\otimes\xi_{d|u_1,\ldots,u_{d-1}},
\end{equation*}
with $\xi_{d|u_1,\ldots,u_{d-1}}$ being a probability measure. For any value of $\Gamma^1_i,\ldots,\Gamma^{d-1}_i$, we suppose that $\Gamma^d_i$ is a random variable with probability measure $\xi_{d|\Gamma^1_i,\ldots,\Gamma^{d-1}_i}$.
Then, the process $(L_t^1,\ldots,L_t^d)_{t\in[0,1]}$ defined by
\begin{equation*}
L_t^j=\sum_{i=1}^{\infty}U_i^{-1}(\Gamma_i^j)\mathbbmss{1}_{[0,t]}(V_i), \quad j=1,\ldots,d
\end{equation*}
is a $d$-dimensional Lévy process $(L_t)_{t\in[0,1]}$ without a Brownian component and drift. The Lévy measure of $L_t$ is $\nu$.
\label{TheoremSimulation}
\end{Theorem}

Proof: The proof is similar to the proof of Tankov \cite{Tankov2005}, Theorem 4.3.\\

In practical simulations, the sum cannot be evaluated up to infinity and one omits very small jumps. The sequence $(\Gamma^1_i)_{i\in \mathbb{N}}$ is therefore only simulated up to a sufficiently large $N,$ resulting in a large value of $\Gamma^1_N$ (see Rosinski \cite{Rosinski2001} for this approximation). Note that large values of $\Gamma^1_i$ correspond to small values of the jumps $U_1^{-1}(\Gamma^1_i)$, since the tail integral is decreasing.

Based on the pair copula construction of the Lévy copula, $\Gamma^2_i,\ldots,\Gamma^d_i$ can be drawn conditionally on $\Gamma^1_i$ in a sequential way. For convenience, assume that the pair Lévy copula has a D-vine structure and that the dimensions are ordered from left to right. The dependence between $\Gamma^1_i$ and $\Gamma^2_i$ is then determined in the first tree of the pair construction by the bivariate Lévy copula $\mathfrak{C}_{1,2}$, and the distribution function $F_{2|\Gamma^1_i}$ of $\Gamma^2_i$ given $\Gamma^1_i$ is derived in the following Proposition.

\begin{Proposition} \emph{} \\
Let $\mathfrak{C}_{1,2}$ be a 2-dimensional Lévy copula with corresponding measure $\mu_{1,2}$. Then, we can decompose
\begin{equation*}
\mu_{1,2}=\mu_1\otimes \xi_{2|u_1},
\end{equation*}
where $\xi_{2|u_1}$ is a probability measure and the distribution function for almost all $u_1\in[0,\infty)$ is given by
\begin{equation*}
F_{2|u_1}(u_2)=\frac{\partial \mathfrak{C}_{1,2}(u_1,u_2)}{\partial u_1}.
\end{equation*}
\label{Proposition2DLevyCopHfunc}
\end{Proposition}

Proof: This is a special case of Tankov \cite{Tankov2005}, Lemma 4.2.

Inverting this distribution function allows the simulation of $\Gamma^2_i.$
Now suppose that we have already simulated the variables $\Gamma^1,\ldots,\Gamma^{d-1}$, $d\geq 3$ and we want to simulate the last variable $\Gamma^d.$
We already know from Theorem \ref{TheoremDisintefration} that the distribution of the last variable, given the first $d-1$, is a specific probability distribution and therefore we are interested in the corresponding distribution function $F_{d|u_1,\ldots,u_{d-1}}$. Having found $F_{d|u_1,\ldots,u_{d-1}}$, we can again invert it and easily simulate a realization of a random variable with this distribution function. The next proposition provides $F_{d|u_1,\ldots,u_{d-1}}$ within the pair construction of the Lévy copula.
\begin{Proposition}  \emph{} \\
Let $d\geq 3$ and $\mathfrak{C}_{1,\ldots,d}$ be a pair Lévy copula, $\mu_{1,\ldots,d}$ the corresponding measure, $\pi$ the projection on the first $d-1$ variables, and $\mu_{1,\ldots,d-1}=\pi\#\mu_{1,\ldots,d}$ the push forward measure. Then, we can decompose
\begin{equation*}
\mu_{1,\ldots,d}=\mu_{1,\ldots,d-1}\otimes\xi_{d|u_1,\ldots,u_{d-1}},
\end{equation*}
where $\xi_{d|u_1,\ldots,u_{d-1}}$ is a probability measure on $\mathbb{R}_+$ with distribution function
\begin{equation*}
F_{d|u_1,\ldots,u_{d-1}}(u_d)=\frac{\partial C_{1,d|2,\dots,d-1}(F_{1|u_2,\ldots,u_{d-1}}(u_1),F_{d|u_2,\ldots,u_{d-1}}(u_d))}{\partial F_{1|u_2,\ldots,u_{d-1}}(u_1)}
\end{equation*}
$\mu_{1,\ldots,d-1}$-almost everywhere. Moreover, $F_{d|u_1,\ldots,u_{d-1}}$ is continuously differentiable.
\label{prop.simugamma}
\end{Proposition}
The proposition is proved in the appendix. Similar to Aas et al. \cite{Aas2009}, it shows how we can iteratively evaluate and invert the distribution function $F_{d|u_1,\ldots,u_{d-1}}$.

\subsection{Maximum Likelihood Estimation}

It is usually not possible to track Lévy processes in continuous time. Therefore, we have to choose a more realistic observation scheme. In the context of inference for pure jump Lévy processes, a common assumption is that it is possible to observe all jumps of the processes larger than a given $\varepsilon$ (see, e.g., Basawa and Brockwell (1978, 1980) \cite{Basawa1978,Basawa1980} or Esmaeili and Kl\"{u}ppelberg \cite{Esmaeili2010}).

Following Esmaeili and Kl\"{u}ppelberg \cite{Esmaeili2011b}, we estimate the marginal Lévy processes separately from the dependence structure.
That is, we use all observations with jumps larger than $\varepsilon$ in a certain dimension and estimate the parameters of the one-dimensional Lévy process.

For the estimation of the dependence structure, i.e., the Lévy copula, we can use the fact that the process consisting of all jumps larger than $\varepsilon$ in all dimensions is a compound Poisson process.
We suppose that all densities $f_1,\ldots,f_d$ of the marginal Lévy measures exist and we denote the parameter vectors of the Lévy copula and the marginal Lévy measures by $\delta, \gamma_1,\ldots,\gamma_d$, respectively.
The likelihood function is given by
\begin{align*}
L^{\varepsilon}(\delta,\gamma_{1}&,\ldots,\gamma_{d})=\\
&e^{-\lambda^{(\varepsilon)}_{1,\ldots,d}t}\prod_{i=1}^{N^{(\varepsilon)}_{1,\ldots,d}}\left[f_{1}(x_{i1},\gamma_1)\cdot \ldots \cdot f_{d}(x_{id},\gamma_d)\mathfrak{c}_{1,\ldots,d}(U_1(x_{i1},\gamma_1),\dots,U_d(x_{id},\gamma_d),\delta) \right],
\end{align*}
where $\lambda^{(\varepsilon)}_{1,\ldots,d}=\mathfrak{C}_{1,\ldots,d}(U_1(\varepsilon,\gamma_1),\dots,U_d(\varepsilon,\gamma_d),\delta)$ , $N^{(\varepsilon)}$ is the number of observed jumps,
and $\mathfrak{c}_{1,\ldots,d}$ is the density of $\mathfrak{C}_{1,\ldots,d}$.
This result also holds for $m$-dimensional marginal Lévy processes with $m<d$ and is stated in Esmaeili and Kl\"{u}ppelberg \cite{Esmaeili2011} for two dimensions.

A straightforward estimation approach would be maximizing the full likelihood function to estimate the dependence structure. This, however, is disadvantageous because of two reasons. The first reason is a numerical one. The likelihood function is not easy to evaluate if more than one parameter is unknown. The second reason is more conceptual. Since we can only use jumps larger than $\varepsilon$ in all $d$ dimensions, we waste a tremendous part of the information about the dependence structure, especially if the dependence structure is weak. For weak dependence structures, the probability that two jumps are both larger than a threshold (conditioned that at least one jump exceeds the threshold) is lower than for strong dependence.

 For both reasons, we estimate the parameters of the bivariate Lévy and distributional copulas of the vine structure sequentially. This is also common for pair copula constructions of distributional copulas (see, e.g., Hob{\ae}k Haff \cite{Haff2012b}). We make use of the estimated marginal parameters and start in the first tree, using all observations larger than $\varepsilon$ in the first and second components to estimate the parameters of $\mathfrak{C}_{1,2}$. We continue this procedure for all other Lévy copulas in the first tree. To estimate the parameter of $C_{1,3|2}$, we use all observations larger than $\varepsilon$ in dimensions one, two, and three, as well as the previously estimated marginal parameters of the first three dimensions and the parameters of $\mathfrak{C}_{1,2}$ and $\mathfrak{C}_{2,3}$. This means that we proceed tree by tree and within the tree, copula by copula or Lévy copula by Lévy copula, respectively. In each step, we make use of the estimated parameters from the preceding steps.

To use the above likelihood for pair Lévy copula constructions, we have to know how to calculate the density $\mathfrak{c}_{1,\ldots,d}$ of a pair Lévy copula.

\begin{Proposition}  \emph{} \\
\label{prop.DensityDecomposition}
Let $\mathfrak{C}_{1,\ldots,d}$ be a pair Lévy copula of the following form
\begin{equation*}
\mathfrak{C}_{1,\ldots,d}(u_1,\ldots,u_d)=\int\limits_{[0,u_2]\times\ldots\times [0,u_{d-1}]}C(F_{1|z_2,\ldots,z_{d-1}}(u_1),F_{d|z_2,\ldots,z_{d-1}}(u_d))d\mu_{2,\ldots, d-1}(z_2,\ldots,z_{d-1})
\end{equation*}
and $\mu_{1,\ldots,d}$ the corresponding measure and suppose that the density $f_{2,\ldots,d}$ of $\mu_{2,\ldots,d-1}$ exists. Then the density of $\mu_{1,\ldots,d}$ exists as well and has the form
\begin{eqnarray*}
f_{1,\ldots,d}(u_1,\ldots,u_d)&=&c(F_{1|u_2,\ldots,u_{d-1}}(u_1),F_{d|u_2,\ldots,u_{d-1}}(u_d))
\label{EquationDensityDecomposition1}\\
&&\cdot \frac{\partial F_{1|u_2,\ldots,u_{d-1}}(u_1)}{\partial u_1} \frac{\partial F_{d|u_2,\ldots,u_{d-1}}(u_d)}{\partial u_d}
\label{EquationDensityDecomposition2}\\
&&\cdot f_{2,\ldots,d}(u_2,\ldots, u_{d-1}).
\label{EquationDensityDecomposition3}
\end{eqnarray*}
\end{Proposition}

This proposition is proved in the appendix and states that we can iteratively decompose the pair Lévy copula into bivariate building blocks and therefore evaluate the density function in an efficient manner.

In contrast to the computation of the density of the pair Lévy copula, it is not easy to evaluate a higher dimensional pair Lévy copula itself. This is no real drawback, since the value of $\mathfrak{C}_{1,\ldots,d}$ is not needed in most cases. For the normalizing constant $\lambda^{(\varepsilon)}_{1,\ldots,d}$ of the likelihood, however, $\mathfrak{C}_{1,\ldots,d}$ has to be evaluated. For this step, we apply Monte Carlo methods. The code may be obtained from the authors on request, so that for convenience, we omit the details here.

%------------------------------------------------------------------------------------------------------------------------
%------------------------------------------------------------------------------------------------------------------------
\section{Simulation Study}
%------------------------------------------------------------------------------------------------------------------------
\label{sec.Simu}

In order to evaluate the estimators, we conduct a simulation study with a $5$-dimensional PLCC. To make the results comparable, all marginal Lévy processes are chosen to be stable Lévy processes with parameters $(\alpha=0.5,\beta=1)$ and all bivariate Lévy copulas in the first tree are Clayton Lévy copulas (see Equation (\ref{glg.clayton})) with parameter $\theta$. The distributional copulas in the higher trees are all Gaussian copulas, i.e.,
   $$C_\rho^{Gauss}(u,v) = \Phi_\rho\left(\Phi^{-1}(u), \Phi^{-1}(v) \right),$$ where $\Phi_\rho$ is the distribution function of the bivariate normal distribution with correlation parameter $\rho$ and $\Phi^{-1}$ the quantile function of the standard normal distribution.

We analyze three different scenarios of dependence structures: high dependence (H), medium dependence (M) and low dependence (L). In scenarios H and M, we choose a D-vine structure of the PLCC, in scenario L a C-vine as it is numerically more appropriate for low dependencies. The D-vine structure refers to a structure where all dimensions in the lowest tree form a line and are each connected to the nearest neighbors, whereas the dimensions in a C-vine structure are connected to only one central dimension (see, e.g., Aas et al. \cite{Aas2009}).
Within a scenario, all Clayton Lévy copulas have the same parameter $\theta$ and all Gaussian copulas have the same parameter $\rho$. The parameter values are summarized in Table \ref{tab.simuParas}.

\begin{table}[htbp]
  \centering
    \begin{tabular}{rrr}
    \toprule
    Scenario & Clayton Parameters $\theta$ & Gaussian Parameters $\rho$ \\
    \midrule
    High dependence (H) & 5     & 0.8 \\
    Medium dependence (M) & 2     & 0.3 \\
    Low dependence (L) & 1     & -0.2 \\
    \bottomrule
    \end{tabular}%
    \caption{Parameters of the PLCC for scenarios H, M and L.
  \label{tab.simuParas}}
\end{table}%

For each scenario, we simulate a realization of a 5-dimensional Lévy process over a time horizon $[0,T]$. We then estimate the parameters of the process from the simulated data using our estimation approach. We choose two different thresholds $\varepsilon=10^{-4}$ and $\varepsilon=10^{-6}$ for jump sizes we can observe, i.e., we neglect jumps smaller than $\varepsilon=10^{-4}$ or $\varepsilon=10^{-6},$ respectively. Each simulation/estimation step is repeated 1000 times. The estimation results are reported in Tables \ref{TableSimuStudyPLCCObserveJumps} and \ref{TableSimuStudyPLCCObserveJumpseM4}. Shown are the true values of the parameters, the mean of the estimates of the 1000 repetitions and resulting estimates for bias and root mean square error (RMSE). Since the parameters in the different trees rely on different numbers of observation (the higher the tree, the more dimensions have to exceed the threshold at the same time) we also report the mean numbers of available jumps per tree.

\begin{figure}[ht]
	\centering
    \includegraphics[width=0.9\textwidth, height=0.6\textwidth]{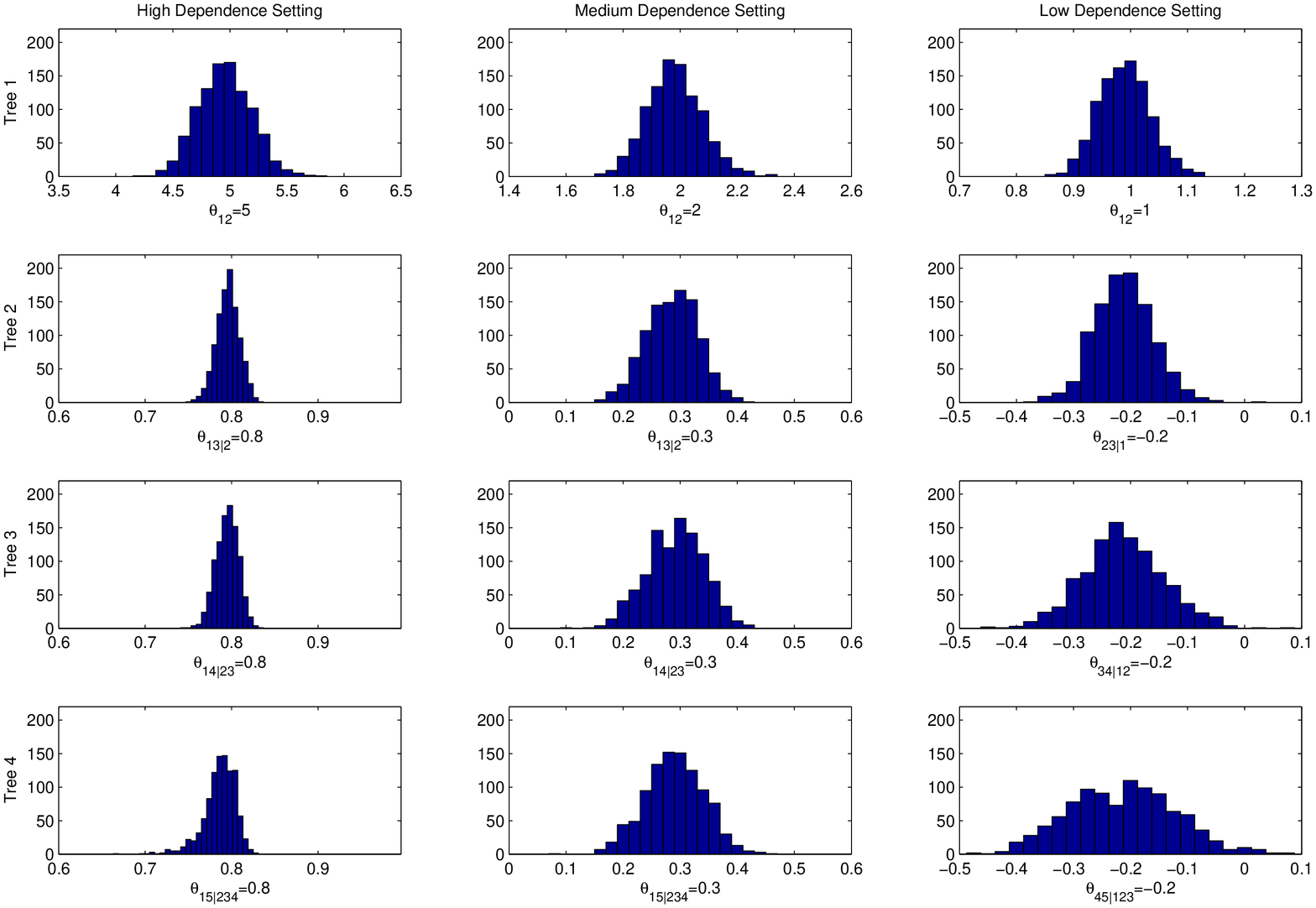}
    \caption{Histograms of the estimation results for a time horizon T=1 and a threshold $\varepsilon = 10^{-6}.$ Each column refers to one scenario, the rows refer to the estimated parameters in the first to fourth tree. \label{fig.schaetzer} }
\end{figure}

Comparing the two tables, we see that the lower threshold leads to a higher number of jumps. We also find that weaker dependence leads to less co-jumps available for the estimation of higher trees than a stronger dependence.
In all cases, the bias is very small. We find, however, that the RMSE is affected by the number of jumps available in certain trees as it increases with decreasing number of jumps. This effect is illustrated in Figure \ref{fig.schaetzer} in terms of histograms of the estimates.

\begin{table}
\begin{center}
\begin{tabular}{lrrrrrr}
\toprule
&Tree& \# Jumps& True Value& Mean &Bias& RMSE\\ \hline
\midrule
High Dep.&1&870.61&5&5.0038&$3.78\cdot 10^{-3}$& $2.33\cdot 10^{-1}$\\
&2&833.51&0.8&0.7987&$-1.28\cdot 10^{-3}$&$1.33\cdot 10^{-2}$\\
&3&814.39&0.8&0.7980&$-1.97\cdot 10^{-3}$&$1.34\cdot 10^{-2}$\\
&4&798.46&0.8&0.7890&$-1.10\cdot 10^{-2}$&$2.19\cdot 10^{-2}$\\
\midrule
Med. Dep.&1&707.18&2&2.0010&$1.02\cdot 10^{-3}$&$9.65\cdot 10^{-2}$\\
&2&573.56&0.3&0.2983&$-1.67\cdot 10^{-3}$&$4.58\cdot 10^{-2}$\\
&3&498.45&0.3&0.2983&$-1.72\cdot 10^{-3}$&$4.97\cdot 10^{-2}$\\
&4&451.69&0.3&0.3001&$1.31\cdot 10^{-4}$&$5.11\cdot 10^{-2}$\\ \midrule
Low Dep.&1&500.10&1&1.0016&$1.63\cdot 10^{-3}$&$4.46\cdot 10^{-2}$\\
&2&267.36&-0.2&-0.1987&$1.31\cdot 10^{-3}$&$4.98\cdot 10^{-2}$\\
&3&163.22&-0.2&-0.1992&$7.96\cdot 10^{-4}$&$7.12\cdot 10^{-2}$\\
&4&113.91&-0.2&-0.2004&$-3.76\cdot 10^{-4}$&$9.50\cdot 10^{-2}$\\
\bottomrule
\end{tabular}
\caption{Results for a time horizon T=1 and a threshold $\varepsilon = 10^{-6}$ for three scenarios from low dependence to high dependence. The columns refer to the number of jumps used in the estimation of parameters within a certain tree, the true value of the parameters, the mean of the estimated parameters, estimated bias and RMSE from 1000 Monte Carlo repetitions. The first three trees contain more than one dependence function and we report the mean values of the estimators in these cases.\label{TableSimuStudyPLCCObserveJumps} }
\end{center}
\end{table}

\begin{table}
\begin{center}
\begin{tabular}{lrrrrrr}
\toprule
&Tree& \# Jumps& True Value& Mean &Bias& RMSE\\ \hline
\midrule
High Dep.&1&87.26&5&5.0403&$4.03\cdot 10^{-2}$&$7.06\cdot 10^{-1}$\\
&2&83.63&0.8&0.7933&$-6.76\cdot 10^{-3}$&$4.61\cdot 10^{-2}$\\
&3&81.69&0.8&0.7810&$-1.90\cdot 10^{-2}$&$5.67\cdot 10^{-2}$\\
&4&80.10&0.8&0.7086&$-9.14\cdot 10^{-2}$&$1.46\cdot 10^{-1}$\\ \midrule
Med. Dep.
&1&70.82&2&2.0312&$3.12\cdot 10^{-2}$&$3.19\cdot 10^{-1}$\\
&2&57.47&0.3&0.2970&$-3.00\cdot 10^{-3}$&$1.50\cdot 10^{-1}$\\
&3&50.00&0.3&0.2844&$-1.56\cdot 10^{-2}$&$1.59\cdot 10^{-1}$\\
&4&45.37&0.3&0.2797&$-2.03\cdot 10^{-2}$&$1.63\cdot 10^{-1}$\\\midrule
Low Dep.
&1&50.21&1&1.0246&$2.46\cdot 10^{-2}$&$1.55\cdot 10^{-1}$\\
&2&26.88&-0.2&-0.2019&$-1.87\cdot 10^{-3}$&$1.67\cdot 10^{-1}$\\
&3&16.42&-0.2&-0.1859&$1.41\cdot 10^{-2}$&$2.57\cdot 10^{-1}$\\
&4&11.47&-0.2&-0.1378&$6.22\cdot 10^{-2}$&$3.44\cdot 10^{-1}$\\
\bottomrule
\end{tabular}
\caption{Results for a time horizon T=1 and a threshold $\varepsilon = 10^{-4}$ for three scenarios from low dependence to high dependence.
The columns refer to the number of jumps used in the estimation of parameters within a certain tree, the true value of the parameters, the mean of the estimated parameters, estimated bias and RMSE from 1000 Monte Carlo repetitions. The first three trees contain more than one dependence function and we report the mean values of the estimators in these cases.
Compared to Table {\protect \ref{TableSimuStudyPLCCObserveJumps}}, the higher threshold $\varepsilon$ results in fewer observed jumps and in higher RMSE of the estimates. \label{TableSimuStudyPLCCObserveJumpseM4}}
\end{center}
\end{table}

%------------------------------------------------------------------------------------------------------------------------
\section{Conclusion}
%------------------------------------------------------------------------------------------------------------------------
\label{sec.Concl}

Lévy copulas determine the dependence of jumps of Lévy processes in a multivariate setting with arbitrary numbers of dimensions. In dimensions larger than two, however, known parametric Lévy copulas are inflexible. In this paper, we develop a multidimensional pair construction of Lévy copulas from 2-dimensional dependence functions which are either Lévy copulas or distributional copulas. The resulting parametric Lévy copula has the desired flexibility, since every regular vine and every bivariate (Lévy) copula can be used in the PLCC. Applications of the concept can be found in operational risk modeling or risk management of insurance companies. In both fields, Lévy copula models have been proposed but their applicability was limited to low-dimensional cases. The pair construction solves these limitations and opens the way to high-dimensional applications. In this paper, simulation and estimation methods are evaluated in a simulation study.

\newpage
%------------------------------------------------------------------------------------------------------------------------
%---------------APPENDIX-------------------------------------------------------------------------------------------------
%------------------------------------------------------------------------------------------------------------------------
\begin{appendix}
\section{Proof of Theorem \ref{TheoremPLCC}}

For the proof of Theorem \ref{TheoremPLCC}, we need a lemma which we state first.

\begin{Lemma} \emph{} \\
\label{LemmaMuMeasurablility}
Let $\mu$ be a positive Radon measure on $\mathbb{R}_+^d$, $f_1:x\mapsto \xi^1_x$ and $f_2:x\mapsto \xi^2_x$ $\mu$-measurable measure-valued maps, where $\xi^1_x$ and $\xi^2_x$ are probability measures on $\mathbb{R}_+$ with corresponding distribution functions $F_x^1$ and $F_x^2$. Let $C$ be a 2-dimensional distributional copula and let $\xi_x^{C}$ be the probability measure defined by the distribution function $C(F_x^1,F_x^2)$ on $\mathbb{R}_+^2$. Then, the map $x\mapsto \xi^C_x$ is $\mu$-measurable.
\end{Lemma}
Proof:
By definition, the maps $x\mapsto \xi^1_x(B_1)$ and $x\mapsto \xi^2_x(B_2)$ are $\mu$-measurable for any $B_1, B_2 \in \mathcal{B}(\mathbb{R}_+)$. This holds in particular for the intervals $[0,b]\in \mathcal{B}(\mathbb{R}_+)$. Therefore, the maps $x\mapsto F^1_x(b_1)$ and $x\mapsto F^2_x(b_2)$ are $\mu$-measurable for any $b_1,b_2 \in \mathbb{R_+}$.
By definition of $\xi_x^C,$ we have
\begin{equation*}
\xi_x^C(B)=C(F_x^1(b_1),F_x^2(b_2))
\end{equation*}
for any rectangle $B\in \{[0,b_1]\times[0,b_2]|b_1,b_2\in\mathbb{R}_+\}$. Since $C$ is a copula, it is continuous and therefore measurable. We get that $x\mapsto \xi_x^C(B)$ is a composition of $\mu$-measurable functions and therefore $\mu$-measurable for any rectangle $B\in \{[0,b]|b\in\mathbb{R}_+^2\}$.
Now that we have shown that $x\mapsto \xi^C_x(B)$ is $\mu$-measurable for any $B\in \{[0,b]|b\in\mathbb{R}_+^2\}$, we use the same argumentation as in the proof of Ambrosio, Fusco, and Pallara \cite{Ambrosio2000}, Proposition 2.6, to show that $x\mapsto \xi^C_x(B)$ is $\mu$-measurable for any $B\in \mathcal{B}(\mathbb{R}^2_+)$. Note that the set of intervals $B\in \{[0,b]|b\in\mathbb{R}_+^2\}$ is closed under finite intersection, it is a generator of the $\sigma$-algebra $\mathcal{B}(\mathbb{R}_+^2)$, and there exists a sequence $(B_h)$ of these intervals such that $\mathbb{R}_+^2=\cup_{h}B_h$.
Denote the family of Borel sets such that $x\mapsto \xi^C_x(B)$ is $\mu$-measurable by $\mathcal{M}$. Obviously, $\mathcal{M}\supset \{[0,b]|b\in\mathbb{R}_+^2\}$. In order to use Ambrosio et al. \cite{Ambrosio2000}, Remark 1.9, we have to show that the following conditions hold:
\begin{enumerate}
\item $(E_h)\in \mathcal{M}$, $E_h\uparrow E \Rightarrow E\in \mathcal{M}$,
\item $E$,$F$, $E\cup F \in \mathcal{M} \Rightarrow E \cap F \in \mathcal{M}$,
\item $E\in \mathcal{M} \Rightarrow \mathbb{R}^2_+\backslash E\in \mathcal{M}$.
\end{enumerate}
 This is already shown in the first part in the proof of Ambrosio, Fusco, and Pallara \cite{Ambrosio2000}, Proposition 2.26.
$\Box$

Now we are able to prove Theorem \ref{TheoremPLCC}.

We show that the integral is well-defined in the first step. From Theorem \ref{TheoremDisintefration} follows that $(u_2,\ldots,u_{d-1})\mapsto \xi_{1|u_2,\ldots,u_{d-1}}$ is $\mu_{2,\ldots,d-1}$-measurable. By the definition of measure-valued maps, $(u_2,\ldots,u_{d-1})\mapsto \xi_{1|u_2,\ldots,u_{d-1}}(B)$ is $\mu_{2,\ldots,d-1}$-measurable for any $B \in \mathcal{B}(\mathbb{R_+})$ and especially for any $B \in \{[0,b]|b\in \mathbb{R_+}\}$. Therefore,
\begin{equation*}
\xi_{1|u_2,\ldots,u_{d-1}}([0,b])=F_{1|u_2,\ldots,u_{d-1}}(b)
\end{equation*}
is $\mu_{2,\ldots,d-1}$-measurable. With the same arguments, we see immediately that $F_{d|u_2,\ldots,u_{d-1}}(b)$ is $\mu_{2,\ldots,d-1}$-measurable for any $b\in\mathbb{R}_+$. Since every copula is continuous, we can use the same arguments as in the proof of Lemma \ref{LemmaMuMeasurablility} to show that
\begin{equation*}
(u_2,\ldots,u_{d-1})\mapsto C(F_{1|u_2,\ldots,u_{d-1}}(u_1),F_{d|u_2,\ldots,u_{d-1}}(u_d))
\end{equation*}
is $\mu_{2,\ldots,d-1}$-measurable and that the integral is well-defined.
To show that $\mathfrak{C}_{1,\ldots,d}$ is indeed a Lévy copula, we have to check the properties of Tankov \cite{Tankov2005}, Definition 3.3.
We start by showing that $\mathfrak{C}_{1,\ldots,d}$ is $d$-increasing. In a first step, we show this property for any $d$-box $B$ where all vertices lie in $\mathbb{R}^d_+$.
For every $(u_2,\ldots,u_{d-1})\in \mathbb{R}^{d-2}_+$ let $\xi_{1,d|u_2,\ldots,u_{d-1}}^C$ be the probability measure on $\mathbb{R}^2_+$ defined by the distribution function $C(F_{1|u_2,\ldots,u_{d-1}}(u_1),F_{d|u_2,\ldots,u_{d-1}}(u_d))$. With Lemma \ref{LemmaMuMeasurablility} we know that $(u_2,\ldots,u_{d-1})\mapsto \xi_{1,d|u_2,\ldots,u_{d-1}}^C$ is $\mu_{2,\ldots,d-1}$-measurable.
By definition of $\mathfrak{C}_{1,\ldots,d}$
\begin{eqnarray*}
\mathfrak{C}_{1,\ldots,d}(u_1,\ldots,u_d)&=&\int\limits_{[0,u_2]\times\ldots \times  [0,u_{d-1}]}C(F_{1|z_2,\ldots,z_{d-1}}(u_1),F_{d|z_2,\ldots,z_{d-1}}(u_d))d\mu_{2,\ldots, d-1}(z_2,\ldots,z_{d-1})\\
&=&\int\limits_{[0,u_2]\times \ldots \times [0,u_{d-1}]}\left(\int\limits_{[0,u_1]\times[0,u_d]}d\xi_u^C\right)d\mu_{2,\ldots,d-1}(z_2,\ldots, z_{d-1})
\end{eqnarray*}
holds, and therefore
\begin{eqnarray*}
\mathfrak{C}_{1,\ldots,d}(u_1,\ldots,u_d) &=& \mu_{2,\ldots,d-1}\otimes\xi_{1,d|u_2,\ldots,u_{d-1}}^C([0,u_1]\times\ldots\times[0,u_d])\\
&=&\mu_{1,\ldots,d}([0,u_1]\times\ldots\times[0,u_d]).
\end{eqnarray*}
Since $\mu_{2,\ldots,d-1}\otimes\xi_{1,d|u_2,\ldots,u_{d-1}}^C$ is a positive and well-defined measure,
\begin{equation*}
V_{\mathfrak{C}_{1,\ldots,d}}(B)=\mu_{2,\ldots,d-1}\otimes \xi_u^C (B)\geq0.
\end{equation*}

In the next step, we denote $u_I:=\{u_i|i\in I\}$ and show that the limit in Equation \eqref{EqPLCCLimes} exists for any $I\subset\{1,\ldots,d\}$ nonempty, $I\neq \{1,\ldots,d\}$. First, suppose that $\{1,d\}\subset I$. Since $I\neq \{1,\ldots,d\},$ we say w.l.o.g. that $\{2\}\notin I$. Since $\mathfrak{C}_{1,\ldots,d}$ is non-decreasing in every component, it suffices to show that
\begin{eqnarray*}
&&\lim_{u_I \rightarrow \infty} \mathfrak{C}_{1,\ldots,d}(u_1,\ldots,u_d)\\&=&\lim_{u_I \rightarrow \infty} \int_{[0,u_2]\times \ldots\times [0,u_{d-1}]}C(F_{1|z_2,\ldots,z_{d-1}}(u_1),F_{d|z_2,\ldots,z_{d-1}}(u_d))d\mu_{2,\ldots,d-1}(z_2,\ldots ,z_{d-1})\\
&=& \lim_{u_{I\backslash\{1,d\}} \rightarrow \infty} \int_{[0,u_2]\times\ldots\times [0,u_{d-1}]}d\mu_{2,\ldots,d-1}(z_2,\ldots,z_{d-1})\\
&=&\lim_{u_{I\backslash\{1,d\}}\rightarrow \infty}\int_{[0,u_2]\times\ldots\times [0,u_{d-1}]}\int_{[0,\infty)}d\xi_{1|z_2,\ldots,z_{d-1}}d\mu_{2,\ldots,d-1}(z_2,\ldots, z_{d-1})\\
&=&\lim_{u_{I\backslash\{1,d\}}\rightarrow \infty}\int_{[0\times\infty)\times[0,u_2]\times\ldots\times [0,u_{d-1}]}d\mu_{1,\ldots,d-1}(z_1,\ldots,z_{d-1})\\
&=&\lim_{u_{I\backslash\{1,d\}}\rightarrow \infty}\mathfrak{C}_{1,\ldots,d-1}(\infty,u_2,\ldots,u_{d-1})\\
&\leq&\mathfrak{C}_{1,\ldots,d-1}(\infty,u_2,\infty,\ldots,\infty)=u_2
\end{eqnarray*}
to prove that the limes exists. We use the dominated convergence theorem (e.g. Ambrosio, Fusco, and Pallara \cite{Ambrosio2000}, Theorem 1.21) and the fact that for every distributional copula $C(u_1,u_2)\leq1$ holds. For the inequality, we use the fact that Assumption \ref{AssumptionLimes} holds for the Lévy copula $\mathfrak{C}_{1,\ldots,d-1}$.
Now, suppose that at least one element of $\{1,d\}$ is not in $I$. W.l.o.g. $\{1\}\notin I$ then we have
\begin{eqnarray*}
&&\lim_{u_I \rightarrow \infty} \mathfrak{C}_{1,\ldots,d}(u_1,\ldots,u_d)\\&=&\lim_{u_I \rightarrow \infty} \int_{[0,u_2]\times\ldots\times [0,u_{d-1}]}C(F_{1|z_2,\ldots,z_{d-1}}(u_1),F_{d|z_2,\ldots,z_{d-1}}(u_d))d\mu_{2,\ldots,d-1}(z_2,\ldots, z_{d-1})\\
&\leq&\lim_{u_{I\backslash\{d\}} \rightarrow \infty} \int_{[0,u_2]\times\ldots\times [0,u_{d-1}]}F_{1|z_2,\ldots,z_{d-1}}(u_1)d\mu_{2,\ldots,d-1}(z_2,\ldots,z_{d-1})\\
&=&\lim_{u_{I\backslash\{d\}} \rightarrow \infty} \int_{[0,u_2]\times\ldots\times [0,u_{d-1}]}\int_{[0,u_1]}d\xi_{1|z_2,\ldots,z_{d-1}} d\mu_{2,\ldots,d-1}(z_2,\ldots,z_{d-1})\\
&=&\lim_{u_{I\backslash\{d\}} \rightarrow \infty} \int_{[0,u_1]\times[0,u_2]\times\ldots\times [0,u_{d-1}]}d\mu_{1,\ldots,d-1}(z_2,\ldots,z_{d-1})\\
&\leq& \mathfrak{C}_{1,\ldots,d-1}(u_1,\infty,\ldots,\infty)=u_1.
\end{eqnarray*}
Now that we have shown that the limit exists, it follows immediately that $\mathfrak{C}_{1,\ldots,d}$ is $d$-increasing on $\overline{\mathbb{R}}^d_+$.
To show that the Lévy copula $\mathfrak{C}_{1,\ldots,d}$ has Lebesgue margins, we can again use the same equations as before and replace ``$\leq$'' by ``='' since in this case $|I|=d-1$ and therefore we can directly use Assumption \ref{AssumptionLimes}. $\Box$

%---------------------------------------------------------------------------------------------------------
\section{Proof of Proposition \ref{prop.simugamma}}

Suppose that $F_{1|u_2,\ldots,u_{d-1}}$ and $F_{d|u_2,\ldots,u_{d-1}}$ are continuously differentiable. For any rectangle $B=([0,u_1]\times\ldots\times[0,u_d]),$ we get by Theorem \ref{TheoremDisintefration}
\begin{equation*}
\int_{\mathbb{R}^{d}_+}\mathbbmss{1}_B(z_1,\ldots,z_d)d\mu_{1,\ldots,d}(z_1,\ldots,z_{d})
=\int_{[0,u_1]\times\ldots\times[0,u_{d-1}]}
F_{d|u_1,\ldots,u_{d-1}}(u_d)
d\mu_{1,\ldots,d-1}(z_1,\ldots,z_{d-1}).
\end{equation*}
By the definition of the pair Lévy copula we see that
\begin{eqnarray*}
&&\int_{\mathbb{R}^{d}_+}\mathbbmss{1}_B(z_1,\ldots,z_d)d\mu_{1,\ldots,d}(z_1,\ldots,z_{d})\\
&=&\int_{\mathbb{R}^{d-2}_+}\left( \int_{\mathbb{R}^2_+} \mathbbmss{1}_B(z_1,\ldots,z_d)d\xi^C_{1,d|u_2,\ldots,u_{d-1}} \right)d\mu_{2,\ldots,d-1}(z_2,\ldots,z_{d-1})\\
&=&\int_{[0,u_2]\times\ldots\times[0,u_{d-1}]}\left(C(F_{1|z_2,\ldots,z_{d-1}}(u_1),F_{d|z_2,\ldots,z_{d-1}}(u_d))\right)d\mu_{2,\ldots,d-1}(z_2,\ldots,z_{d-1})\\
&=&\int_{[0,u_2]\times\ldots\times[0,u_{d-1}]}\\
&&\bigg(
\int_{[0,u_1]}\frac{\partial C(F_{1|z_2,\ldots,z_{d-1}}(z_1),F_{d|z_2,\ldots,z_{d-1}}(u_d))}{\partial F_{1|z_2,\ldots,z_{d-1}}(z_1)}
\frac{\partial F_{1|z_2,\ldots,z_{d-1}}(z_1)}{\partial z_1}dz_1\bigg)
d\mu_{2,\ldots,d-1}(z_2,\ldots,z_{d-1})\\
&=&\int_{[0,u_2]\times\ldots\times[0,u_{d-1}]}\\
&&\bigg(
\int_{[0,u_1]}\frac{\partial C(F_{1|z_2,\ldots,z_{d-1}}(z_1),F_{d|z_2,\ldots,z_{d-1}}(u_d))}{\partial F_{1|z_2,\ldots,z_{d-1}}(z_1)} d\xi_{1|z_2,\ldots,z_{d-1}}(z_1)
\bigg)d\mu_{2,\ldots,d-1}(z_2,\ldots,z_{d-1})\\
&=&\int_{[0,u_1]\times\ldots\times[0,u_{d-1}]}
\frac{\partial C(F_{1|z_2,\ldots,z_{d-1}}(z_1),F_{d|z_2,\ldots,z_{d-1}}(u_d))}{\partial F_{1|z_2,\ldots,z_{d-1}}(z_1)}
d\mu_{1,\ldots,d-1}(z_1,\ldots,z_{d-1}),
\end{eqnarray*}
and therefore
\begin{equation*}
F_{d|u_1,\ldots,u_{d-1}}(u_d)=\frac{\partial C(F_{1|u_2,\ldots,u_{d-1}}(u_1),F_{d|u_2,\ldots,u_{d-1}}(u_d))}{\partial F_{1|u_2,\ldots,u_{d-1}}(u_1)}
\end{equation*}
holds $\mu_{1,\ldots,d-1}$-almost everywhere. The fact that this result does not only hold for fixed values of $u_d$ but for all $u_d\in \mathbb{R}_+$ is already shown in the proof of Tankov \cite{Tankov2005}, Lemma 4.2.
Since $F_{1|u_2,\ldots,u_{d-1}}$, $F_{d|u_2,\ldots,u_{d-1}}$ are continuously differentiable and C is by Assumption \ref{AssumptionContDiff} also continuously differentiable, we get immediately that $F_{d|u_1,\ldots,u_{d-1}}$ is differentiable and
\begin{equation*}
\frac{\partial F_{d|u_1,\ldots,u_{d-1}}(u_d)}{\partial u_d}=\frac{\partial^2 C(F_{1|u_2,\ldots,u_{d-1}}(u_1),F_{d|u_2,\ldots,u_{d-1}}(u_d))}{\partial F_{1|u_2,\ldots,u_{d-1}}(u_1)\partial F_{d|u_2,\ldots,u_{d-1}}(u_d)}\frac{\partial F_{d|u_2,\ldots,u_{d-1}}(u_d)}{\partial u_d}
\end{equation*}
is a composition of continuous functions and therefore continuous. Finally, all bivariate Lévy copulas are by Assumption \ref{AssumptionContDiff} continuously differentiable and therefore, the proposition follows by complete induction. $\Box$

\section{Proof of Proposition \ref{prop.DensityDecomposition}}

This statement follows from the definition of the pair Lévy copula construction, since
\begin{eqnarray*}
\mathfrak{C}_{1,\ldots,d}(u_1,\ldots,u_d)&=&\int\limits_{[0,u_2]\times\ldots\times [0,u_{d-1}]}C(F_{1|z_2,\ldots,z_{d-1}}(u_1),F_{d|z_2,\ldots,z_{d-1}}(u_d))d\mu_{2,\ldots, d-1}(z_2,\ldots,z_{d-1})\\
&=&\int_{[0,u_2]\times\ldots\times [0,u_{d-1}]}\bigg(\int_{[0,u_1]\times[0,u_d]}c(F_{1|z_2,\ldots,z_{d-1}}(z_1),F_{d|z_2,\ldots,z_{d-1}}(z_d))\\
&& \frac{\partial F_{1|z_2,\ldots,z_{d-1}}(z_1)}{\partial z_1} \frac{\partial F_{d|z_2,\ldots,z_{d-1}}(z_d)}{\partial z_d}d(z_1,z_d)\bigg)d\mu_{2,\ldots,d-1}(z_2,\ldots,z_{d-1})\\
&=&\int_{[0,u_2]\times\ldots\times [0,u_{d-1}]}\bigg(\int_{[0,u_1]\times[0,u_d]}c(F_{1|z_2,\ldots,z_{d-1}}(z_1),F_{d|z_2,\ldots,z_{d-1}}(z_d))\\
&& \frac{\partial F_{1|z_2,\ldots,z_{d-1}}(z_1)}{\partial z_1} \frac{\partial F_{d|z_2,\ldots,z_{d-1}}(z_d)}{\partial z_d}d(z_1,z_d)\bigg)\\
&&f_{2,\ldots,d-1}(z_2,\ldots,z_{d-1})d(z_2,\ldots,z_{d-1})\\
&=&\int_{[0,u_1]\times\ldots\times[0,u_d]}
c(F_{1|z_2,\ldots,z_{d-1}}(z_1),F_{d|z_2,\ldots,z_{d-1}}(z_d))\\
&& \frac{\partial F_{1|z_2,\ldots,z_{d-1}}(z_1)}{\partial z_1} \frac{\partial F_{d|z_2,\ldots,z_{d-1}}(z_d)}{\partial z_d}f_{2,\ldots,d-1}(z_2,\ldots,z_{d-1})
d(z_1,\ldots,z_d)
\end{eqnarray*}
as stated. $\Box$

\end{appendix}

\newpage

%\bibliographystyle{plain}
%\bibliography{Literatur}

\end{document}